\documentclass{ws-procs975x65}

\begin{document}
\vskip -2.0cm
\title{Properties of accretion shock waves in viscous 
flows with cooling effects}

\author{Santabrata Das$^1$ and Sandip K. Chakrabarti$^{2,3}$}

\address{$^1$ARCSEC, Sejong University,
         98 Gunja-Dong, Gwangjin-Gu,
         Seoul 143-747, South Korea.
E-mail: sbdas@canopus.cnu.ac.kr\\}

\address{$^2$S. N. Bose National centre for Basic Sciences,
  JD-Block, Sector III,\\
  Salt Lake, Kolkata, 700 098, India. E-mail:chakraba@bose.res.in\\
  $^3$ Centre for Space Physics, Chalantika 43, Garia Station Rd.,
  Kolkata 700084, India.}

\begin{abstract}
We study the properties of the shock waves for a viscous accretion flow
having low angular momentum in presence of synchrotron cooling. We
present all possible accretion solutions in terms of flow parameters.
We identify the region of the parameter space for steady and oscillating 
shocks and show the effect of various energy dissipation processes
on it. We discuss the role of the shock waves while explaining the 
observations from black hole candidates.
\end{abstract}

\keywords{accretion, accretion disc -- black hole physics--shock waves.}

\bodymatter

\section{Introduction}

The shock induced accretion flow is currently one of the most promising 
self-consistent hydrodynamical accretion disk model, since it explains
the spectral states of black holes as well as quasi-periodic oscillations
(QPOs) most satisfactorily (\cite{c90,c96} and references therein).  
Since then, various groups of workers extensively studied the properties of the shock 
waves (e.g.,  \cite{yk95,ft04}) in different astrophysical contexts. A fully 
self-consistent solutions of isothermal, viscous transonic flows was obtained by
Chakrabarti \cite{c90} considering sub-Keplerian flow at the inner part of the disc. In a rotating 
accretion flow, centrifugal force acts as a barrier \cite{c96,cd04,d07}
that triggers the flow to undergo shock transition at a location where the Rankine-Hugoniot shock
conditions (hereafter RHCs) are satisfied. However, so far, the study of 
viscous flows in presence of Synchrotron cooling has not been done.
In the present paper, we concentrate on the study of a stationary, 
axisymmetric, viscous accretion solutions around a Schwarzschild black 
hole in presence of Synchrotron cooling. The space-time geometry 
around a non-rotating black hole is approximated by the pseudo-Newtonian potential \cite{pw80}. 
We study all the relevant dynamical flow variables in terms of the inflow parameters.
We identify the solution topologies which are essential for shock formation.
The effects of viscosity and cooling are expected to be different
while deciding the dynamical structure of the accretion flow since cooling reduces 
the flow energy while viscosity not only tends to heat the flow but transports the 
angular momentum from inner edge to the outer edge. We find that shock waves, 
standing or oscillating, can form even at a very high dissipation limit. The hot and 
dense post shock flow is the natural site of the hot radiation in the accretion disc 
and is believed to be a powerful tool in understanding the spectral
properties of black holes \cite{ct95}, QPOs of the hard X-rays \cite{msc96,rcm97} and the 
formation of the accretion-powered relativistic bipolar outflows/jets 
\cite{dc07}. In this paper, we discuss these issues. 

\section{Results and Discussions}
\def\figsubcap#1{\par\noindent\centering\footnotesize(#1)}
\begin{figure}[b]%
\begin{center}
\vskip -0.4 in
\hskip -0.9 in
  \parbox{1.0in}{\epsfig{figure=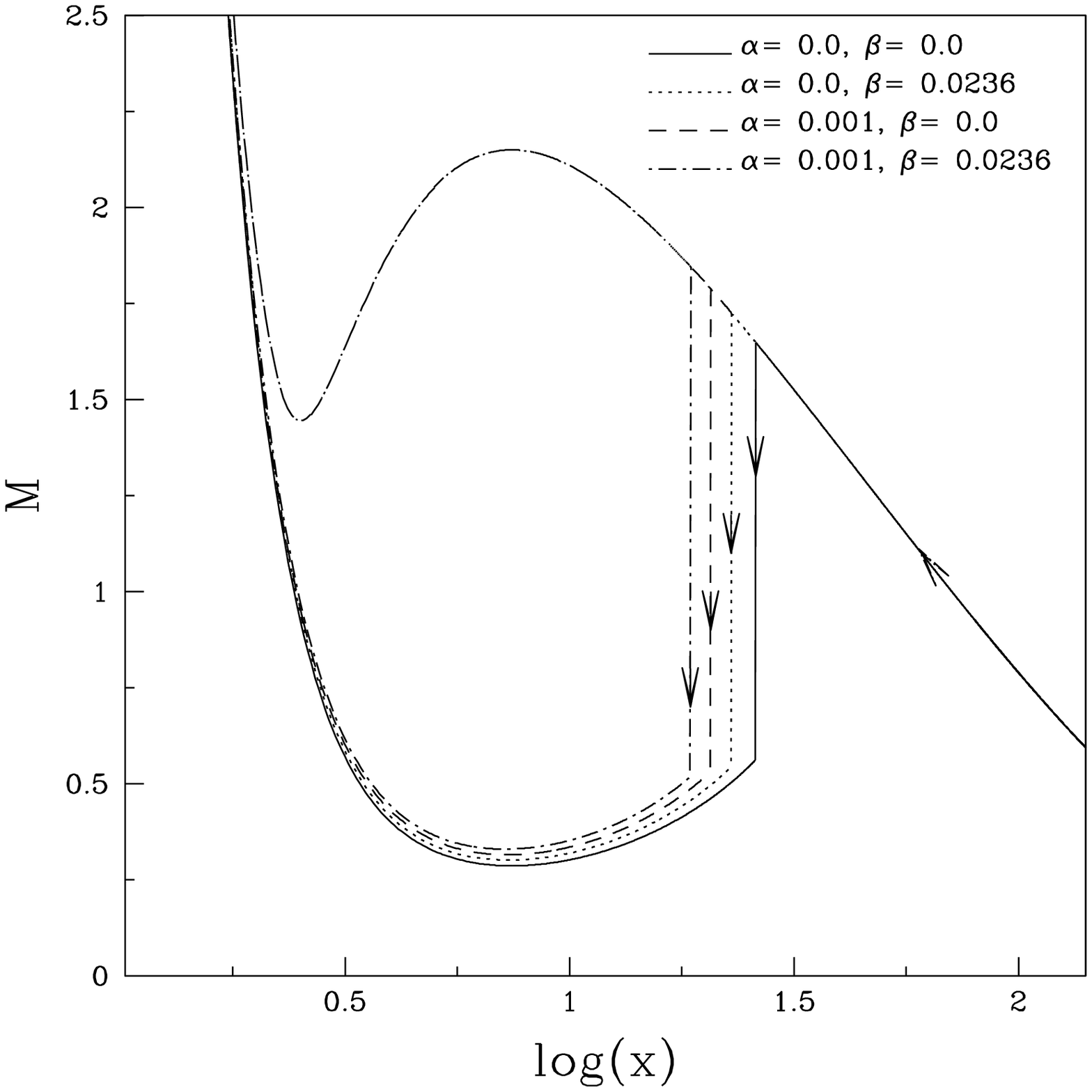,width=2in}\figsubcap{a}}
\hskip 0.7 in
  \parbox{1.0in}{\epsfig{figure=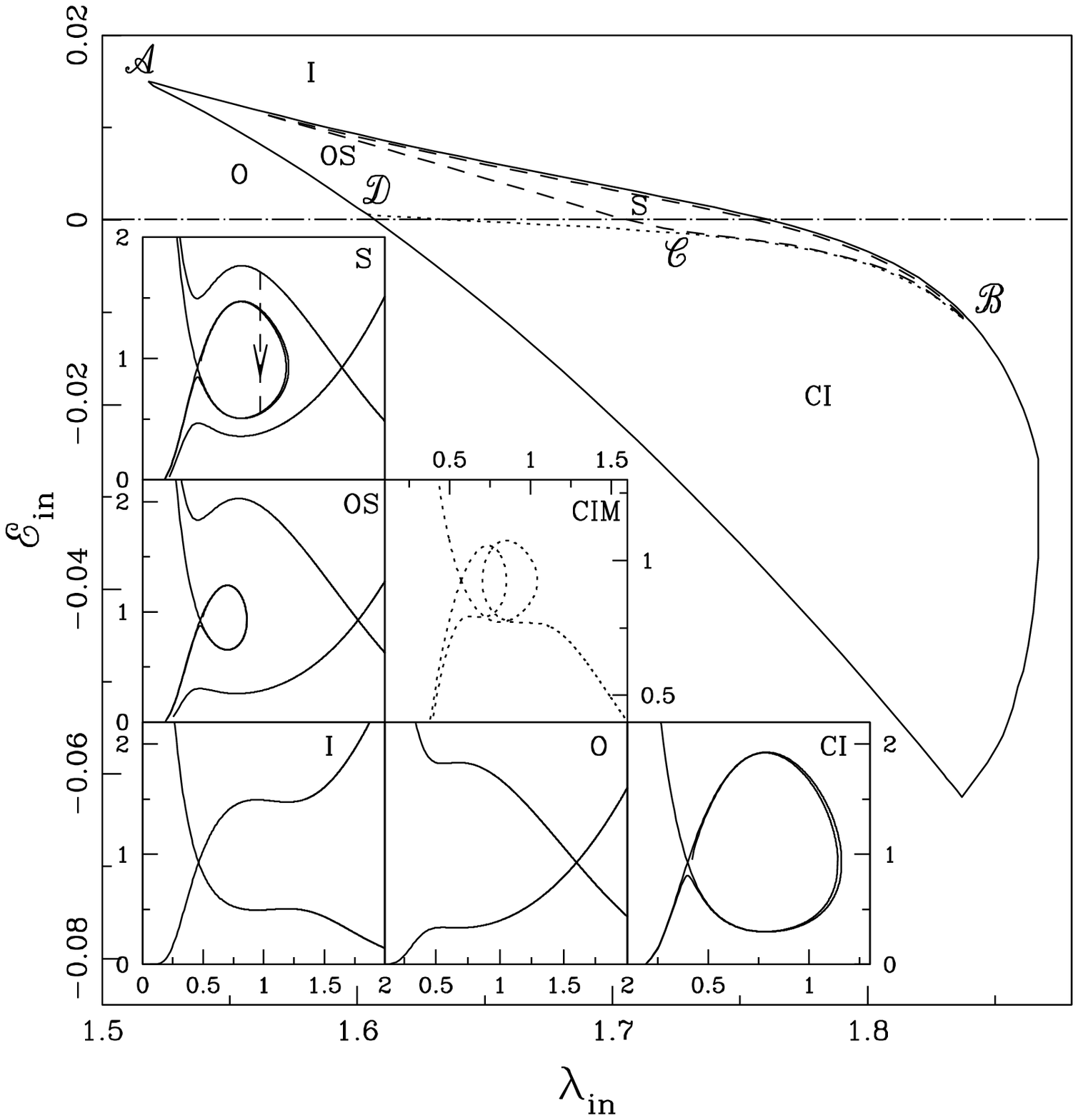,width=2in}\figsubcap{b}}
\hskip 0.7 in
  \parbox{1.0in}{\epsfig{figure=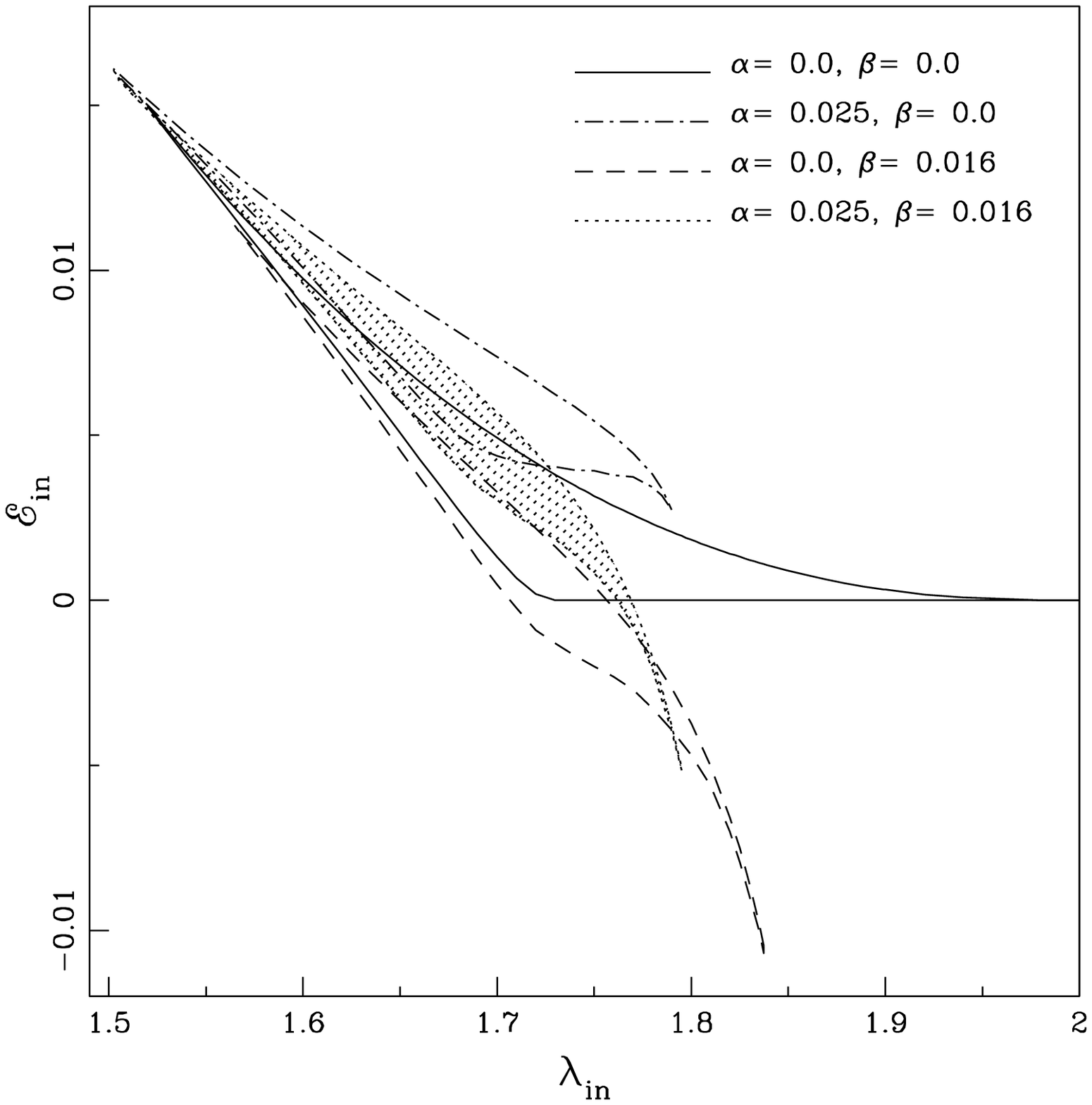,width=2in}\figsubcap{c}}
  \caption{(a) Variation of Mach number with the logarithmic radial distance. 
           Dissipation parameters and corresponding shock locations are
           marked. (b) Sub-division of parameter space in the 
           ${\mathcal E}_{in}-\lambda_{in}$ plane according to the nature of
           solutions. (c) Variation of the parameter space for standing 
           shocks for different dissipation parameters (marked).}%
  \label{fig1.2}
\end{center}
\end{figure}

A set of classical shock solutions are presented in Fig. 1a in terms
of viscosity ($\alpha$) and/or cooling parameter ($\beta$). 
We inject matter sub-sonically at the outer edge of the disc $x_{inj}=145$ 
with local energy $\mathcal {E}_{inj} = 3.3663 \times 10^{-3}$
and angular momentum $\lambda_{inj} = 1.725$. 
Solid vertical line represents the shock location for non-dissipative
($\alpha =0$ and $\beta = 0$) flow. When $\alpha \ne 0$, the shock front 
moves inward depicted by the dashed vertical line. In an accretion flow, 
viscosity transports angular momentum outwards causing the 
reduction of the centrifugal pressure and at the same time, the viscous heating increases
flow energy. Since the shock moves forward as $\lambda(x)$ get reduced, 
we can conclude that the centrifugal force is the primary cause for shock formation.
When $\beta \ne 0$, the shock location again proceeds towards the horizon.
In the hot and dense post-shock flow cooling is more 
efficient which reduces the post-shock pressure causing the shock to move forward further.
When both $\alpha \ne 0$ and $\beta \ne 0$, shock
location is predicted at $x_s = 18.62$ and is indicated by vertical 
dot-dashed line. Here, the shock front is shifted significantly
due to the combined effects of viscosity and cooling. 

In Fig. 1b, we separate the regions of the parameter space spanned by the specific energy 
(${\mathcal E}_{in}$) and specific angular momentum ($\lambda_{in}$) at the inner sonic 
point $(x_{ci})$ according to the flow topologies for $\beta = 0.00787$ in inviscid limit. Solid 
boundary represents the region for closed topologies passing through 
$x_{ci}$. At the inset, all possible solutions [Mach number
vs. log(x)] with parameters chosen from different region of the parameter 
space are presented. The box S represents the shock solution. 
The box OS shows an accretion solution having oscillating 
shock. For higher cooling, we obtain a new solution topology as shown 
in box CIM. We draw this solution with dotted curve as it is obtained 
for higher cooling parameter. The solution I passes directly through 
the $x_{ci}$ before entering into the black hole.
The solution O represents a flow which passes
through the outer sonic point $(x_{co})$ only. The box CI shows a closed flow solution
passing through $x_{ci}$. This type of solution does not extend to the 
outer edge to join smoothly with any Keplerian disc and flow is expected to 
be unstable. 

In Fig. 1c, we compair solutions with
different dissipation parameters. Solid boundary is the parameter
space for standing shock for $\alpha = 0$ and $\beta = 0$. 
For $\alpha \ne 0$, the effective region of the parameter space
for standing shock separated by dot-dashed curve shrinks and moves  
towards the higher ${\mathcal E}_{in}$ and the lower $\lambda_{in}$ 
regime \cite{cd04}. When $\beta \ne 0$, the parameter space is  reduced
and shifts to the lower energy sides \cite{d07}. For flows with $\alpha \ne 0$ and $\beta \ne 0$, 
the parameter space for the shock settles down at an intermediate region
(shaded part). This shows that the viscosity and the synchrotron cooling
act in opposite directions in deciding the parameter space.

\section{Conclusion}

We studied the properties of viscous accretion flow around a 
non-rotating BH in presence of synchrotron cooling. 
We found that the flow can have shock waves
even when the viscosity and synchrotron cooling are high.
The standing shocks form closer to the BH 
when dissipation is increased and these shocks are centrifugal pressure supported.
We obtained the parameter space for the standing shocks for
various dissipation parameters and showed that the effective region of
the parameter space shrinks as dissipation rises.
The parameter space
for oscillating shock where RHCs are not satisfied but the flow is
likely to pass through a shock has been identified. 
We also pointed out that the viscosity and the cooling have opposite effects
in deciding the parameter space for stationary shock waves.
Moreover, since the shocks form closer to the BH, QPO frequency
increases with the enhancement of accretion rate as  
observed in several BH candidates.

\vskip 0.1cm
\noindent{Acknowledgments:}
SD is thankful for financial support to KOSEF through ARCSEC, Korea.

\end{document}